\documentclass[%
 reprint,
 amsmath,amssymb,
 aps,
]{revtex4-2}
\usepackage{graphicx}% Include figure files
\usepackage{dcolumn}% Align table columns on decimal point
\usepackage{bm}% bold math
\usepackage{color}
\usepackage[normalem]{ulem}
\usepackage{hyperref}
\usepackage{aas_macros}

\def\g{$\gamma$-ray}

\def\lum{\rm erg~s$^{-1}$~}
\def\frm{\emph{Fermi}}

\begin{document}

\title[Gamma-ray halo...]{Detection of  gamma-ray halos around nearby late-type galaxies}% Force line breaks with \\

\author{M.~S.~Pshirkov}
 \email{pshirkov@sai.msu.ru}
\affiliation{
 Sternberg Astronomical Institute, Lomonosov Moscow State University, Universitetsky prospekt 13, 119992, Moscow, Russia\\
}%
\affiliation{
 Institute for Nuclear Research of Russian Academy of Sciences, 60th October Anniversary Prospect, 7a,
117312, Moscow, Russia\\
}%

\author{B.~A.~Nizamov}%
 \email{nizamov@physics.msu.ru}
\affiliation{%
 Sternberg Astronomical Institute, Lomonosov Moscow State University,\\ Universitetsky prospekt 13, 119992, Moscow, Russia\\
}%

\date{\today}% It is always \today, today,

\begin{abstract}
Various  theoretical models predict the existence of extended \g {} halo around normal galaxies that could be produced by interactions of  cosmic rays with the circumgalactic medium or by annihilation or decay of hypothetical dark matter particles. Observations of M31, the closest massive galaxy, also corroborate this possibility.  
%One of possible ways to shed light on the nature of dark matter (DM) is the search for gamma-ray emission due to its decay or annihilation in circumgalactic halos. However, gamma-ray observations of normal galaxies are rather scarce.
%only a handful of local normal galaxies have been detected in this energy range. Moreover, because of a moderate angular resolution of modern instruments, only one galaxy apart from the Milky Way can be firmly claimed as an extended gamma-ray source, which is the Andromeda galaxy.
    In this study we search for gamma-ray emission from the galaxies within 15~Mpc at energies higher than 2 GeV and try to assess its spatial extension. We use the latest catalog of local galaxies and apply a simple yet robust method of aperture photometry. By imposing the mass, energy, and spatial cuts, we selected a set of 16 late-type galaxies  and found a statistically significant excess  above the background level: a p-value of $3.7\times10^{-7}$ at $E>2$ GeV, reaching maximal significance of $p-\mathrm{val}=2.3\times10^{-8}$ for a subset of front-converted events with $E>2$~GeV, where the angular resolution is higher. More importantly, our analysis shows that this excess can be ascribed to an extended source with a radius $\sim 0.3^\circ$ rather than a point-like one. This, for $D=15$~Mpc, corresponds to a physical halo radius of $r_h=80$~kpc.  In contrast, 6 early-type galaxies, which satisfied the same cuts, showed no excess. Our results are supported by the stacking likelihood analysis technique which significantly ($5.6\sigma$) detected an extended excess. The difference between the late- and early-type galaxies and a rather irregular shape of the extended source that we found, indicate that this high-energy emission is more likely caused by the interactions of cosmic rays with the circumgalactic medium, in preference to DM annihilation/decay  processes.  %\maxim{reformulate DM part along Ref 2 remarks}.
%\begin{description}
%\end{description}
\end{abstract}

\maketitle

%\tableofcontents

%%%%%%%%%%%%%%  INTRO  %%%%%%%%%%%%%%%%%%%%%%%%%%
\section{\label{sec:intro}Introduction}
At the present level of the instrument sensitivity the high-energy $\gamma$-ray sky ($E>100$~MeV) is dominated by the active galactic nuclei (AGNs) of various types: AGNs account for 4008 of the 7194 total sources in the latest catalog of the Fermi-LAT sources \cite{4FGL-DR4}.  %Much more modest contribution comes from star forming galaxies, due to their lower intrinsic luminosities, especially for normal galaxies without  ongoing bursts  of star-formation. 
Only a handful of  galaxies without AGNs have been detected \cite{Fermi2012a,Ajello2020, Ambrosone2024} as individual sources, although by some estimations the whole population of such galaxies could be a major contributor to the extragalactic diffuse \g {} background \cite{Roth2021}.
The $\gamma$-ray emission in star-forming galaxies mostly originates in collisions of galactic cosmic rays (CRs) with the interstellar medium. 
Due to the limited angular resolution of the instruments (around $1^{\circ}$ at GeV energies),  almost  all  detected galaxies  are best described as point sources.
Spatially extended emission was detected around two normal galaxies: first, the famous Fermi bubbles were detected in our Galaxy \cite{FB1,FB2}% --large,  $\mathcal{O}(10~\mathrm{kpc})$ lobes above and below the Galactic center
; second, there is growing certainty that M31 is also surrounded by an  extended \g {} halo \cite{Pshirkov2016,Karwin2019}.
Extended emission could arise  from  previous  phases of AGN activity, similarly to the cases of  the $\gamma$-ray lobes of the Cen~A and Fornax~A galaxies \cite{CenA, FornaxA},  or it could be produced by the population of  galactic CRs, gradually leaking into the circumgalactic medium and accumulating there on time scales of Gyrs \cite{Crocker2011, Feldmann2013}. Alternatively, \g {} halo could emerge from processes of annihilation or decay of dark matter (DM) particles  \cite{Bringmann2012}.   Both observations of M31 and theoretical models predict that MW-like galaxies have  \g  {} luminosity in $10^{38}-10^{39}~\mathrm{erg~s^{-1}}$ range.

%As individual sources are expected to be  weak and lie below the detection  threshold, we analyzed aggregated observations of nearby massive galaxies and stacked sources in order to increase our sensitivity.  
In this \emph{Letter} we report the discovery of a statistically significant  ($p\mathrm{-val}=3.7\times10^{-7}$) excess of photons at energies $>2$~GeV around local ($D<15$~Mpc) massive late-type galaxies. 
Observations at higher energies, where the \frm-LAT angular resolution is considerably better, allowed us to show that the excess is extended, with a radius  $\sim 0.3^{\circ}$ .
% The \emph{Letter} is organized as follows:  in  Section~\ref{sec:met} we describe the data and method of data analysis, Section~\ref{sec:dis} contains our results and discussion, and we draw conclusions in Section~\ref{sec:conclusions}.
%%%%%%%%%%%%%%%%%%%%%%%%%%%%%%%%%%%%%%%%%%%%%%%%%%%

%%%%%%%%%%%%%%  DATA & METHODS  %%%%%%%%%%%%%%%%%%%%%
%\maxim{XXX}
\section{\label{sec:met}Data and methods}
For our analysis, we used the photon database of \frm-LAT \cite{Atwood2009}. We selected photons belonging to the \emph{SOURCE}  class (reconstruction version PASS8R3) and with energies $E>1$~GeV and zenith angle $\theta<105^{\circ}$. %in order to avoid contamination from the Earth's limb. Observations span the time interval from 04 Aug 2008 (MET=239557417) to 09 Aug 2024 (MET=744865715). 
Observations span the time interval from  Aug 2008 to  Aug 2024. For the  analysis with the Fermi Science Tools (version  2.2.0), we adopted standard quality cuts.
We utilized one of the latest catalogs of nearby galaxies 
\cite{Ohlson2024}. There are mass, distance, and morphological type estimates for 15424 galaxies closer than 50 Mpc. 
Our target set was constructed using several cuts: on the galaxy mass, its distance, galactic latitude and its  sky position relative to \frm{} sources.
First, we estimated the distance threshold, demanding that the observed number of photons in 16 years of the observations for a source with a luminosity $L(>1~\mathrm{GeV})\sim5\times 10^{38}~$\lum {} was larger than 10. This yielded an upper distance limit of $D_{max}=15$~Mpc. The lower limit comes from a requirement that the sought halo, which has a radius  in 50--100 kpc range, should not be too extended, i.e., have an angular size greater than $1^{\circ}$, which sets a lower distance limit of $D_{min}=5$~Mpc. 
The   mass threshold  was set at $M_*=10^{10}~M_{\odot}$, where $M_{*}$ is the stellar mass of the galaxy. % Ohlson et al. derived the stellar masses from the color -- mass-to-light ratio from \cite{Taylor2011} where $g-i$ photometry is available, and they extended this ratio to other colors using galaxies for which a $g-i$ color is available along with some other color [$(g-r)_0$, $(B-V)_0$, $(B-R)_0$].
%One should bear in mind that mass estimates from color--$M/L$ ratios can suffer systematic uncertainties. For example, Ohlson et al. compared the masses they obtained with those from dynamical estimates from \cite{Cappellari2011}. The latter appeared to be $\approx 0.3-0.5$~dex larger than those of Ohlson et al. The authors suggest that the discrepancy could be due to the particular initial mass function they assumed.  MAYBE OMIT DISCREPANCIES
Adoption of mass and distance cuts reduced the number of sources to 89. Adoption of a higher mass threshold, e.g., $10^{10.5}M_{\odot}$, would  result in a much smaller number of available targets. A  lower  threshold would lead to a significant decrease in the expected signal if it were produced in CR-related models, but a much smaller decrease in DM-related models.

We employed a simple statistical method similar to aperture photometry: we compared the number of photons observed within a circle of radius $R$ centered on the source (the ON-region) with the number of photons expected in this region in absence of any source. The latter was calculated from the number of photons observed in the OFF-region -- the annulus with the inner and outer radii of $R$ and $2R$, respectively. %In most cases, $R$ was taken to be $PSF_{68}$, i.e. the 68\% containment angle of the point-spread function, but in some cases we also tried $R = 1.5PSF_{68}$ and $R = 3PSF_{68}$. 
The smallest angular scale that can be effectively probed with that method corresponds to $R=PSF_{68}$, i.e. the 68\% containment angle of the point-spread function,  which  depends on the energy and the conversion type. %(See Table \ref{tab:e_psf}).% We tried several lower energy thresholds which is detailed lower.

Due to the low expected signal level, we selected galaxy targets located far from potential interfering sources and regions of high background. Thus, we imposed a latitude cut, removing all low-latitude galaxies with $|b|<20^{\circ}$. Even a weak, nearby \g {} source could lead to a spurious detection or an overestimation of the background in the OFF-region, which  would greatly diminish the sensitivity of our method. Hence, we removed from the sample all the galaxies with a neighboring 4FGL-DR4 source closer than  $3PSF_{68}$. These are the sources from the Data Release 4 of the fourth full catalog of LAT sources; this release  is based on 14 years of survey data. Additionally, in order to  eliminate targets with  close  \g {} sources not included in the catalog, we performed a standard  likelihood analysis with \textit{fermipy} package and removed seven galaxies: NGC~3368, NGC~3379, NGC~3384, NGC~3675, NGC~4818, NGC~5055, and NGC~5248 from our set as well (see below). Finally, we set a threshold energy $E_0$: on the one hand, we would like to have it as low as possible to increase  photon statistics; on the other hand,  PSF size quickly increases for lower energies, making our last cut prohibitively restrictive. We chose $E_0=2~$GeV, $PSF_{68}(E_0)$=0.5$^{\circ}$. 
After  implementing latitude and proximity cuts we were left  with 22 sources out of 89 initial sources that satisfied both mass and distance conditions. 

For our selected threshold of $R=0.5^{\circ}$, there are 1872 events in 22 ON-regions, while the expected number estimated from the OFF-regions is 1728 events, and the corresponding p-value of $1.3\times10^{-4}$. 
However, when we split our sample into subsets of early-type (6 galaxies) and late-type galaxies (16 galaxies), a striking difference emerges. The early-type galaxies show a deficit of events (480 observed vs. 513 expected). In contrast, the late-type galaxies show a significant excess (1392 observed vs. 1215 expected).

Consequently, we focused on this subset (see Table 1 in the Supplemental Material \cite{Supplemental}).

 We  repeated our analysis for different energy cuts and different conversion types --  both front+back and front-converting only for an increased angular resolution (see Table 2 in the Supplemental Material \cite{Supplemental}). Our results are presented in  Table \ref{tab:res}.

One should keep in mind that trying different thresholds reduces statistical significance: we investigated spatial extent of the signal and its spectral shape, using higher energy thresholds and different conversion types. Also, in order to estimate the spatial extent of the source, we naturally had to try several values of ON-region radius, which again affects the statistical significance.  However, calculating a precise numerical value for this correction is difficult, as the trials are highly non-independent. We do not, therefore, make a
correction for trials in the statistical significances quoted below.

\begin{table*}
\caption{\label{tab:res} 
Results of our analysis for different energy thresholds, conversion types and radii of the ON-region $R$.  
For each combination of energy threshold $E_0$, conversion type, and ON-region radius $R$, we list the observed/expected number of events and the corresponding p-value (assuming Poisson background). Bolded cells indicate  $R\approx \mathrm{PSF}_{68}(E, \text{conversion type})$} %We have not performed our analysis for radii smaller than $PSF_{68}$.}
\begin{ruledtabular}
\begin{tabular}{lccc}
  &	$0.25^{\circ}$ &	$0.35^{\circ}$ &	$0.5^{\circ}$  \\
\hline \\
2 GeV       &	-                       & 713/607 $1.50\times10^{-5}$                    & \textbf{1392/1215}, $\mathbf{3.66\times10^{-7}}$ \\
2 GeV front & -                         & \textbf{391/312}, $\mathbf{9.17\times10^{-6}}$ & {758/617}, {$2.30\times10^{-8} $}  \\
3 GeV       & -                         & \textbf{392/326}, $\mathbf{2.12\times10^{-4}}$ & {772/670}, {$6.29\times10^{-5} $} \\
3 GeV front & \textbf{116/106, ~0.177}  & {226/165}, {$3.89\times10^{-6} $}              & {435/340}, {$4.32\times10^{-7   } $} \\
5 GeV       & \textbf{95/93, ~0.431}    & {195/153}, {$6.16\times10^{-4} $}              & {373/324}, {$4.14\times10^{-3} $} \\
5 GeV front &  57/54, ~0.359            & {118/79}, {$2.52\times10^{-5} $}               & {220/168}, {$7.09\times10^{-5} $} \\
10 GeV      &  43/36, ~0.140            & {84/55}, {$1.66\times10^{-4} $}                & {150/124}, {0.01} \\
10 GeV  front     &  27/20, ~0.077            & {51/28}, {$6.02\times10^{-5} $}                & {88/68}, {0.01} \\

30 GeV      & 4/7.3, ---             & {15/12}, 0.228                                 & {26/24}, {0.368} \\
\end{tabular}
\end{ruledtabular}
\end{table*}

The maximal signal was achieved for front-converted photons with  $E_0=2$~GeV and $R=0.5^{\circ}$: 758 observed photons vs. 617 expected photons,  $\mathrm{p-value}=2.3\times10^{-8}$. 
%%%%%%%%%%%%%%%%%%%%%%%%%%%%%%%%%%%%%%%%%%%%%%%%%%%%%%%%

%%%%%%%%%  RESULTS & DISCUSSION  %%%%%%%%%%%%%%%%%%%%%%%
\section{\label{sec:dis}Results and discussion}
Could our detected excess arise because of some flaw in our adopted approach? Although we did not observe any excess signal around six early-type galaxies, we performed additional tests with a larger number of targets to settle the issue.  We repeated our analysis in the neighboring mass bin $10^9~M_{\odot}<M<10^{10}~M_{\odot}$. There are 199 galaxies in the 5--15 Mpc distance range, and after imposing spatial cuts we were left with 51 galaxies satisfying our conditions, 14 early- and 37 late-type galaxies. No significant excess was detected: 4243 photons were observed versus 4284 expected for the entire sample, and 3048 versus 3130 for the late-type subset. Thus, a possible DM-related contribution does not appear to be the major cause of the excess observed for $M>10^{10}~M_{\odot}$ late-type galaxies.

Gamma-ray signal could also be produced by the interaction of CR with the matter of the galactic disk. We calculated the expected signal from the galaxies in our set using the relation between gamma-ray and infrared luminosities from \cite{Fermi2012a} and infrared data from the IRAS catalog \cite{IRAS}. The estimated gamma-ray luminosities appeared to be on the order $10^{38}~\mathrm{erg~s}^{-1}$, i.e. an order of magnitude smaller than the signal we found (see below).

%\bulat{Orphan paragraph -->}We can conclude that the possible DM-related contribution is not the major cause of the excess observed for $M>10^{10}~M_{\odot}$ late-type galaxies. 

\subsection{Morphology}
  Our results show that the excess was not produced by  point-like sources, such as  low-luminosity AGNs in the centers of galaxies or galaxies themselves. If this were the case, the maximal significance at different energies  would be achieved at angular scales corresponding to PSF size at those energies, which is manifestly not the case (see Table \ref{tab:res}). 
  The angular size corresponding to the maximal signal decreases  with improving angular resolution: from $\sim0.5^{\circ}$ at 2 GeV to $\sim0.3^{\circ}$ at 3~GeV. However, it does not decrease further for energies up to 10~GeV, where the angular resolution is twice as high.
  These  results are consistent with a  spatial excess with a radius of  $\sim0.3^{\circ}$, which, for $D=15$~Mpc, corresponds to a physical halo radius of $r_h=80$~kpc.

\subsection{Likelihood analysis and test statistics}
\begin{figure}%[b]
\includegraphics[scale=0.9]{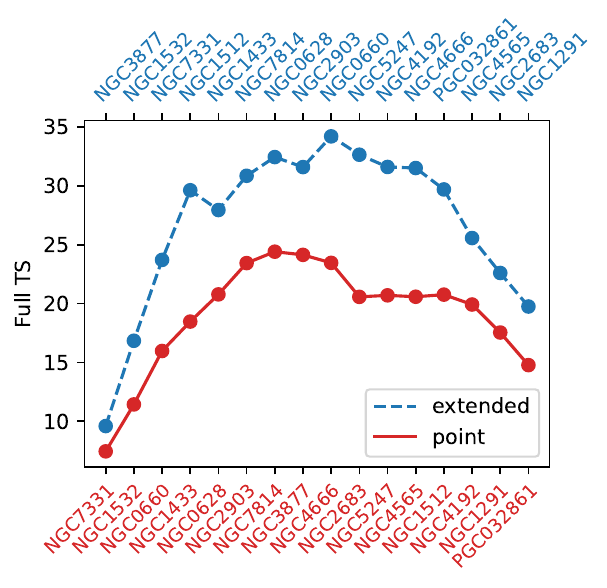}% Here is how to import EPS art
\caption{\label{fig:evolution} 
Cumulative test statistic (TS) for the stacking analysis at $E_0 = 2$~GeV, showing the results for a point-source model (\textit{solid red line}) and an extended source model (\textit{dashed blue line}).}
\end{figure}

We also analyzed our set, using standard Fermi Science tools \footnote{https://github.com/fermi-lat}, such as \emph{gtlike}, which employs the maximum likelihood approach \cite{Mattox1996}. Even the most prominent candidate, NGC~0628, was detected at 2 GeV only at the test statistic level of  $\text{TS}\sim6$  ($\sim2.4\sigma$)  when we included it either as a point source in our source model or an extended uniform disk with a $0.3^{\circ}$ radius. %Comparable level of detection was achieved with an extended uniform disk templates with a radius in a 0.2-0.4$^{\circ}$ range, although as it is evident from  Fig. \ref{fig:cmaps} such a simple template could not effectively reproduce actual pattern of emission.  At the highest energies, $E>10$~GeV point source model fared worse, giving $\text{TS}\sim2$ and a \textit{decrease} in total likelihood, $\Delta LLH=-0.6$, while an extended uniform 0.3 degree disk model provided a slight \textit{increase}, $\Delta LLH=2.4$.  These values are marginally significant, only hinting on a prevalence of an extended template.
We performed a stacking analysis using the code \textsc{fermi\_stacking}\footnote{\url{https://fermi-stacking-analysis.readthedocs.io/en/latest/}}. %which has been previously applied in different studies, e.g., in a search for dark matter annihilation in the Milky Way dwarf spheroidal galaxies \cite{McDaniel2024}.
The pipeline is described in \cite{Paliya2019}: for each individual source, the code assigns it a flux value and a spectral index value from the given intervals and computes its TS for every combination of these values. This gives a 2D-array of TS values. Such arrays for all the sources under study are then added to produce the final TS array such as the one shown in Fig.~\ref{fig:karwin}. The flux and index values corresponding to the global maximum can be regarded as average values for the studied sample. %We performed the stacking analysis for several $E_0$ and, by a slight modification of the code, for a point and extended test source (for the latter, the RadialDisk model was used with the radius of $0.3^\circ$). 
The code  produces an evolution plot, which shows the maximum stacked TS versus the number of sources in  the stack. Each point corresponds to the global TS value after the corresponding source is added to the stack. The plot for the analysis with $E_0=2$~GeV is shown in Fig.~\ref{fig:evolution} for point source and extended radial disk models.
%The lower $x$-axis and the red solid line correspond to stacking of point sources, while the upper $x$-axis and the blue dashed line correspond to stacking of extended sources. The sources on each axis are sorted in descending order of individual TS's. For example, in the point source analysis, the source NGC~7331 has the largest individual TS, and the first red point corresponds to the stack consisting of NGC~7331 only. The second-largest TS is obtained for NGC~1532, and the second red point corresponds to a stack consisting of NGC~7331 and NGC~1532, and so on. Note that the positions of maxima in individual TS arrays can be different from the position of the global maximum. In fact, an individual TS for some source at the position of the global maximum can be negative which is why the evolution plots have a decreasing part. Note also that the individual TS-ranking is not the same in the point and extended source analysis. For example, NGC~7331 has the largest TS in the point source analysis, but only the third-largest one in the extended source analysis.
From Fig.~\ref{fig:evolution}, it is evident that the extended stacked source achieves a considerably larger TS value than the point source.

\textsc{fermi\_stacking} uses the \textit{fermipy} library to search for new sources not present in 4FGL. Among the late-type galaxies which passed our cuts, five sources (NGC~3368, NGC~3675, NGC~4818, NGC~5055, and NGC~5248) appeared to have neighbors within $1.5^\circ$, and we excluded them as in the case of 4FGL source proximity.

We split our set of of late-type galaxies into two subsets of seven and nine sources, respectively.
 In the first one we included sources giving positive TS contribution (NGC0628, NGC0660, NGC1512, NGC1532, NGC3877, NGC7331, NGC7814) and the other contained the rest (NGC1291, NGC1433, NGC2683, NGC2903, NGC4192, NGC4565, NGC4666, NGC5247, PGC032861). Not surprisingly, the first subset performed  much better at 2 GeV (710 observed vs 558 expected, $p\mathrm{-value}=3.7\times10^{-10}$) compared to the second one (682 vs 658, $p\mathrm{-value}=0.18$). 
 However, the situation is different at higher energies; for example, at 10 GeV and $R=0.35^{\circ}$ , the contributions are closer for the first subset (41 vs. 23, $p\mathrm{-value}=4.5\times10^{-4}$ ) and the second (43 vs. 32, $p\mathrm{-value}=3.6\times10^{-2}$).
  This behavior shows that the properties of our sources are far from being identical -- they have  different luminosities and different spectral indices, and the subset of the weaker sources demonstrates a much harder spectrum.
The spread in the properties of individual sources and the impossibility of describing their emission with a single, identical template would lead the stacking approach that we used to underestimate the true statistical significance of the extended emission detection.

\subsection{Spectrum and power}
\begin{figure}
\includegraphics[scale=0.6]{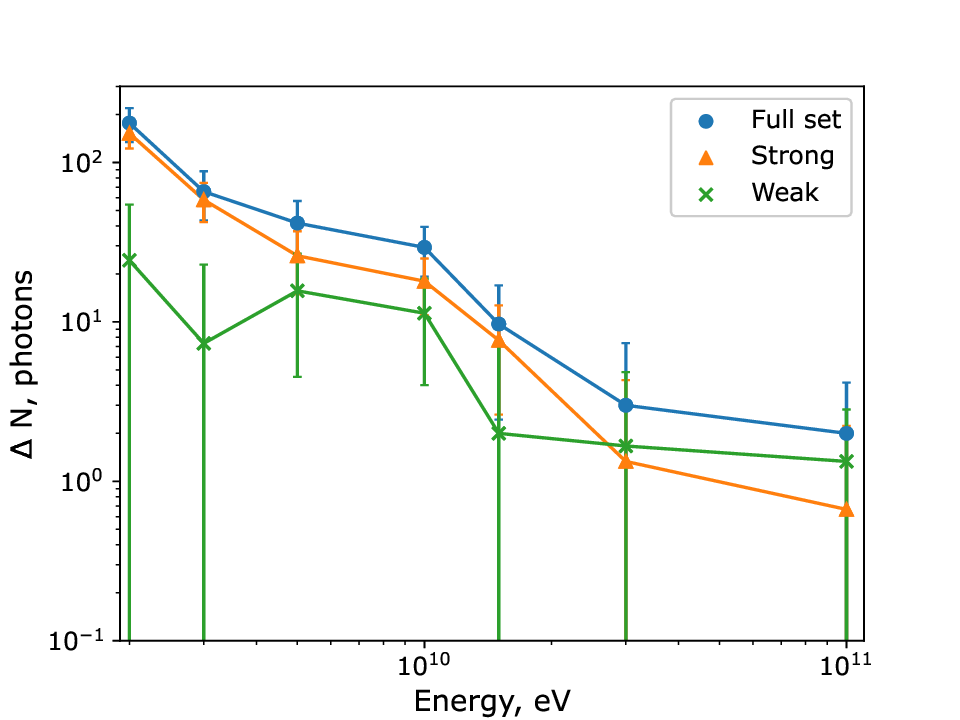}% Here is how to import EPS art
\caption{\label{fig:spectrum_full} 
Integrated spectra for the full sample (16 galaxies) and the 'strong' and 'weak' subsamples. The quantity $\Delta N(E)$ is the total number of excess photons with energy greater than $E$.}
\end{figure}

%We could not easily obtain mean luminosity and characteristic spectral index of the sources due to an obvious heterogeneity  of our set. 
Because of the obvious heterogeneity of the set, we separately analyzed two subsamples ('weak' and 'strong') defined above. We estimated indices and luminosity very roughly, using the number counts of excess photons and assuming that the excess had a characteristic $\sim0.3^\circ$ size.
Results for the full set and both subsets are presented in Fig. \ref{fig:spectrum_full} and Supplemental Material Fig. 1 \cite{Supplemental}, where it can be seen that the galaxies in the 'weak' subset have a low-energy cut-off and a hard spectrum ($\alpha_{\mathrm{weak}}\sim1.8$ at $E > 5$~GeV). In contrast, the 'strong' subset exhibits spectral behavior across the entire 2–200 GeV range that is consistent with a single power law, with a spectral index of $\alpha_{\mathrm{strong}}\sim2.4$.
%Both values should be taken as crude estimates only: the aperture photometry approach has rather limited accuracy for this task and even after dividing the set into 'weak' and 'strong' parts the resulting subsets are not very homogenous. Encouragingly, the analysis with \textsc{fermi\_stacking} gives very close values of spectral indices: $2.8^{+0.6}_{-0.4}$ and $1.7^{+1.1}_{-0.5}$ for the 'strong' and 'weak' subsets respectively.
The estimates of average luminosities in the 2-200 GeV energy range, obtained using catalog distances and total \frm-LAT exposure from 2008-2024, are comparable for galaxies in both subsamples, $L_{\mathrm{strong}}\sim5.3 \times10^{39~}$\lum and $L_{\mathrm{weak}}\sim1.8 \times10^{39~}$\lum. 

%%%%%%%%%%%%%%%%%%%%%%%%%%%%%%%%%%%%%%%%%%%%%%%%%%%%%%

\section{\label{sec:conclusions}Conclusions}
In this paper, we searched for \g{} signal from nearby galaxies ($D<15$~Mpc) using  aperture photometry. We extracted  a sample of galaxies with stellar masses larger than $10^{10}~M_{\odot}$ from the up-to-date catalog \cite{Ohlson2024} and selected all the sources  with galactic latitudes $|b|>20^{\circ}$ residing further than $1.5^{\circ}$ from \g{} sources from \frm-LAT 4FGL-DR4 catalog and from additional point sources (with $TS >16$) discovered in a dedicated likelihood analysis. We detected a statistically significant signal at energies higher than 2 GeV from the  set of 16 late-type galaxies, with a $p-\mathrm{value}\sim3.7\times10^{-7}$. The highest significance was achieved for front-converted events at these energies, with a  $p-\mathrm{value}\sim2.3\times10^{-8}$.
Analysis at different energies showed that the excess is spatially extended, with an angular size of $\sim0.3^{\circ}$, corresponding to a linear size of   $\sim80$~kpc   at a distance of 15 Mpc. The \g{} properties of 
selected late-type  galaxies are not identical: seven out of 16 galaxies have  a higher average luminosity and demonstrate a soft spectrum with a spectral index $\alpha\sim2.4$, while the rest are less luminous and have  a  harder spectrum, with $\alpha\sim1.8$ and  a low-energy cut-off at energies $E<5$~GeV.

%In Table~\ref{tab:galaxies} we show the sample of late-type galaxies where the extended gamma-ray halos were discovered.

%In Table~\ref{tab:e_psf} PSF values are shown for the relevant photon energies.

%In Table~\ref{tab:res} the results of the aperture photometry analysis are summarized.

%The \g {} emission from star forming galaxies can contribute to the isotropic \g {} background as suggested in \cite{Feldmann2013}. We plan to investigate this contribution in a forthcoming paper.

\begin{acknowledgments}
The authors thank Prof. Igor Moskalenko for fruitful discussions. We thank the anonymous referees for their constructive criticisms and suggestions that  significantly enhanced the quality of this work.
The work of the authors was supported by the Ministry of Science and Higher Education of Russian Federation
under the contract 075-15-2024-541 in the framework of the Large Scientific Projects program within the national project "Science". This research has made use of NASA’s Astrophysics Data System. %The numerical part of the work was done at the computer cluster of the Theoretical Division of INR RAS.
\end{acknowledgments}
%comment out the following line for journal submission
%uncomment the following line for journal submission
\providecommand{\noopsort}[1]{}\providecommand{\singleletter}[1]{#1}%

\newpage
\appendix*
%\section{Tables}

\section{End Matter}
\textit{Aperture photometry caveats. -- }
The estimates of the statistical significance are conservative at the lower energies, 2 and 3 GeV (front+back conversions), where $R\sim PSF_{68}$. In our approach we assumed that all the excess photons were contained inside $R$, and the OFF-regions were free from them. However, it is certainly not the case  at lower energies with $R\sim PSF_{68}$, where a significant fraction ($\sim1/3$) of the signal photons spill into the OFF region, artificially raising the background estimate and diluting the detection significance.  A correction for this spill-over effect yields an expected 1182 events (compared to 1215), which would reduce the p-value to $1.5\times10^{-9}$.
 However, for the sake of consistency, we do not quote these values in  Table \ref{tab:res}.

Could our cut on neighboring \frm{} sources  (further than $3PSF_{68}(2~\mathrm{GeV})=1.5^{\circ}$) be too mild? Is the observed excess merely due to pollution from strong, nearby sources that, by chance, are located near our targets?  We performed our analysis for several directions  at a distance of  1.5 degrees  (3 PSF at 2 GeV) from the strongest high-latitude ($|b|>20^{\circ}$)  source, PSR~J1836+5925 (4FGL~J1836.2+5925). Instead of an excess,  we found  a strong deficit,  $\sim310$ observed vs $\sim480$ expected. We can safely state that with our adopted cuts strong sources can only boost the number of events in OFF-regions and consequently create an  apparent \emph{deficit} in the central regions. Additionally, we have checked that even a very strong source at 1.5 degrees distance does not affect our estimates of the background at energies higher than 2 GeV because of  the rapidly shrinking PSF size.

\begin{figure}
\includegraphics[scale=0.8]{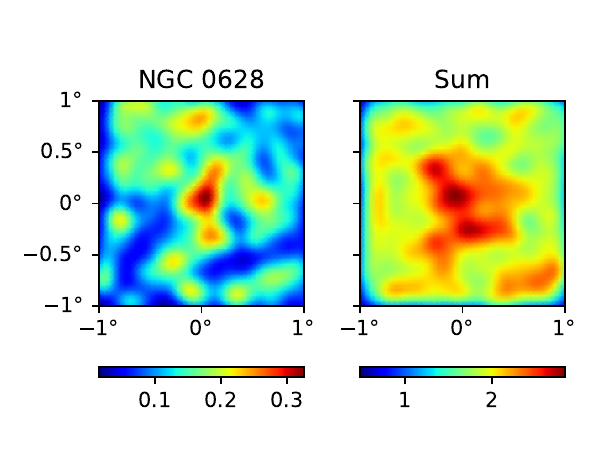}% Here is how to import EPS art
\caption{\label{fig:cmaps} 
Count maps for \textit{(left)} NGC 0628 and \textit{(right)} the stack of all 16 sources, using front-converted photons with $E > 2$~GeV. Maps are smoothed with a Gaussian kernel ($\sigma=0.1^{\circ}$). Coordinates are given relative to the map center.}
\end{figure}

\textit{Origin of the signal. -- } At energies higher than 2 GeV for front converted photons, we constructed joint count map from the count maps of 16  individual sources  (see Fig. \ref{fig:cmaps}).
The excess is clearly visible and has an irregular shape: this fact disfavors  a simple model of a smooth halo, like one of decaying DM. Instead it looks like a superposition of weaker  substructures from the individual sources: as an illustration we show in the same figure the count map of the strongest individual source, NGC 0628 -- there is a hint of lobe-like structures protruding from the central part of the galaxy.

If the excess were related to DM, we would most likely observe such signal for early-type, high-mass galaxies as well, which is not the case. Second, we would expect considerable signal from the lower-mass set ($10^9~M_{\odot}<M<10^{10}~M_{\odot}$): in this mass range there is only weak dependence of the expected halo mass on the stellar mass (e.g. \cite{Shankar2006}): a decrease in the stellar mass from $10^{10}$ to $10^{9}~M_{\odot}$ leads only to corresponding decrease of 0.4 dex in the mass of the halo.
In this narrow mass range we could roughly expect linear dependence of DM-related signal on the mass of the halo. 
Therefore, from our results for $E_0=2$~GeV and $R=PSF_{68}$, we would expect an excess of around 170 events from the 37 late-type galaxies, instead of the 80-photon deficit that we actually observe.

\begin{figure}[b]
\includegraphics[scale=0.9]{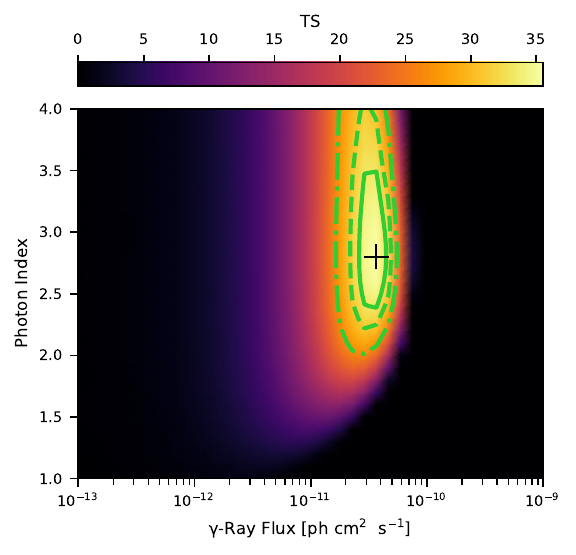}% Here is how to import EPS art
\caption{\label{fig:karwin} 
Stacked TS array for the seven sources that provide positive contributions to the total TS in the extended source analysis with $E_0 = 2$~GeV. The green lines denote the 1$\sigma$, 2$\sigma$, and 3$\sigma$ confidence levels. The black cross indicates the location of the maximum TS value.}
\end{figure}

\textit{Notes on likelihood analysis. -- } The somewhat lower detection significance obtained from the likelihood analysis with   \textsc{fermi\_stacking} can be ascribed to natural limitations of this method. Likelihood analysis is model-dependent, hence its results depend on the spatial and spectral shape of the sources. We showed in the previous subsection that the spatial shape of our sources is irregular, while in our stacking analysis we used a circular template. In contrast, aperture photometry does not rely on the specific spatial distribution of the photons within the ON-region. The same holds in the spectral domain: while likelihood analysis specifies the spectral shape of the sources, aperture photometry only deals with integral photon excess above the chosen threshold energy. There are also quite clear drawbacks to the stacking implementation that we applied: as the fluxes of all sources are assumed to be identical and equal to the average flux, for a weaker source it would lead to a  worse fit and accordingly to a decrease in the aggregate TS. However, almost all these weaker sources give a coherent positive contribution to the statistical significance in our ON-OFF approach and are detected there as well, albeit with a lower significance. In Fig.~\ref{fig:karwin}, we show the stacked TS array for extended source analysis where only sources with positive contributions to the global TS are added to the stack.
The maximum global TS is 35, the best flux is $3.67^{+0.98}_{-1.38} \times 10^{-11}$~ph~cm$^{-2}$~s$^{-1}$, and the best index is $2.8^{+0.7}_{-0.5}$.

\end{document}